\documentclass[
    ,final            
	,mathptm
  ]
  {aipproc}

\layoutstyle{6x9}

\begin{document}

\title[Dust Attenuation caused by Isothermal Turbulent Media]{Dust Attenuation caused by 
	Isothermal Turbulent Media}

\author{J\"org Fischera}{
  address={Research School of Astronomy and Astrophysics,
Institute of Advanced Studies, The Australian National University,
Cotter Road, Weston Creek, ACT 2611 Australia},
  ,email={fischera@mso.anu.edu.au}
}


\begin{abstract}
	Our ability to correct the observational photometry of galaxies depends upon
	our knowledge of the attenuation of light produced by the dust contained in the
	interstellar medium.
	We will present a model based on the statistical properties of the local density in isothermal
	turbulence which might be appropriate to calculate the radiative transport through
	the diffuse interstellar medium. The model will be applied to study the attenuation
	caused by a distant foreground screen and a non scattering slab where the sources
	are mixed within the turbulent medium. It will be shown how the turbulent structure affects
	the attenuation curve and how the attenuation curve varies with the thickness
	of a foreground screen.
\end{abstract}

\maketitle


\section{Introduction}
It is a well known phenomena that the collective star light from galaxies, in particular at UV and optical wavelengths, is obscured by dust grains distributed in the diffuse interstellar medium. To derive accurate measurements of important parameters such as the star-formation rate as a function of redshift, the famous Madau-plot \citep{Madau1996}, a correction can be essential but is still uncertain.

For sake of clarity we
want to differentiate between the dust \emph{extinction} in the case of point sources such as stars and the \emph{attenuation} in the case of extended sources such
as galaxies. In this scheme the dust \emph{extinction} is caused by dust absorption and dust scattering along the line of sight and is simply proportional to the wavelength dependent extinction coefficient $k_\lambda$ and the column density $N$. In contrast the dust \emph{attenuation} is the wavelength dependent loss of the intrinsic light of all stars and can be explained by an effective optical depth. This optical depth is determined by several different conditions such as the relative distribution of emitting stars and the attenuating dust, the viewing angle to the galaxy, the variation of dust properties with radiation field and density throughout the ISM, and the structure of the ISM itself, which is far from homogeneous. In addition it might be important that light scattered out of the observed direction might be partly compensated by photons scattered into the observed direction. It is no surprise that, due to its complexity, the attenuation of the starlight of galaxies is still a rather unsolved problem.

\section{Model}

Many different attempts have been made to take the inhomogeneous nature of the ISM of galaxies into account either to model the spectral energy distributions \citep{Popescu2000, Gordon2001} or to derive corrections for dust attenuation \citep{Witt2000, Tuffs2004, Pierini2005}. 
The model presented in the following is based on a realistic description
of an inhomogeneous structure of the local density caused by turbulent motion.
The basic idea of this model is the fact that the transmission of a distant foreground screen is, if we neglect any variations of the optical properties, primarily determined by the probability distribution function (PDF) of the
column density. By using a simplified description of the turbulent density structure this PDF
can be related in a simple way to its statistical properties and the thickness $\Delta$ of the dusty screen in front of the emitting sources.

\subsection{Model of the Isothermal Turbulent Medium}

The turbulent medium is taken to be isotropic and we assume that inside a certain interval
the power spectrum $P(\mbox{\boldmath$k$})$=$|\rho(\mbox{\boldmath$k$})|^2$ 
of the local density $\rho(\mbox{\boldmath$r$})$
can be described by a simple power law $\tilde P(k)\propto k^{n}$ with power $n$. The scaling relation extends from a minimum scale $L_{\rm min}$ to a maximum scale $L_{\rm max}$. 

As has been shown by using hydrodynamic simulations of turbulence in compressible fluids
the PDF of the local density can be described by a log-normal density distribution if the turbulence is approximately isothermal \citep{Vazquez1994, Padoan1997, Passot1998}. As found analytically \citep{Nordlund1999}, the log-normal density distribution should be the exact solution if supersonic isothermal turbulence is considered.
The PDF of the local density $\rho$ in the ISM is therefore approximated by:
\begin{eqnarray} 
  p(\rho)&=&\frac{1}{\rho}\,p(\ln\rho)=\frac{1}{\sqrt{2\pi}\sigma\rho}\,e^{\frac{1}{2}x^2/\sigma_{\ln\rho}^2}\\
  \mbox{with}~x&=&\ln\rho-\ln\rho_0
\end{eqnarray}
where $\sigma_{\ln\rho}$ is the standard deviation of the log-normal density distribution and where $\rho_0$ is defined by the mean value $\ln\left<\rho\right>=\ln\rho_0 +{0.5\sigma^2_{\ln\rho}}$.

In general the density contrast increases as 
consequence of higher compression resulting from higher Mach numbers $M$.
In particular for non-magnetised forced turbulence it has been found  
by using 3D-simulation \citep{Padoan1997, Nordlund1999} that the standard deviation  
\begin{equation}
   \sigma_\rho = \left<\rho\right>\sqrt{e^{-\frac{1}{2}\sigma_{\ln\rho}^2}-1}
\end{equation}
of the local density is almost linearly correlated with the Mach number by
\begin{equation}
   \label{densmachcorr}
   \sigma_{\rho}\approx \beta M\left<\rho\right>
\end{equation}
where $\beta\approx 0.5$. The more general case appears to be more complicated as
for magnetised turbulence no simple correlation between density contrast and
Mach number has been found. As might be also expected from the 
additional pressure support to the plasma provided by magnetic fields, it seems
that the density contrast becomes weaker when magnetised
turbulence is considered \citep{Nordlund1999, Ostriker2001}.

The variance of the local density is related to the power spectrum of the turbulent density structure by:
\begin{equation}
	\sigma_\rho^2 = \frac{1}{(2\pi)^3}\int {\rm d}\mbox{\boldmath$k$}\,\tilde P(\mbox{\boldmath$k$}).
\end{equation}

\subsection{PDF of the Column Density}

\paragraph{The variance of the column densities}
\begin{figure}
  \includegraphics[width=0.6\hsize]{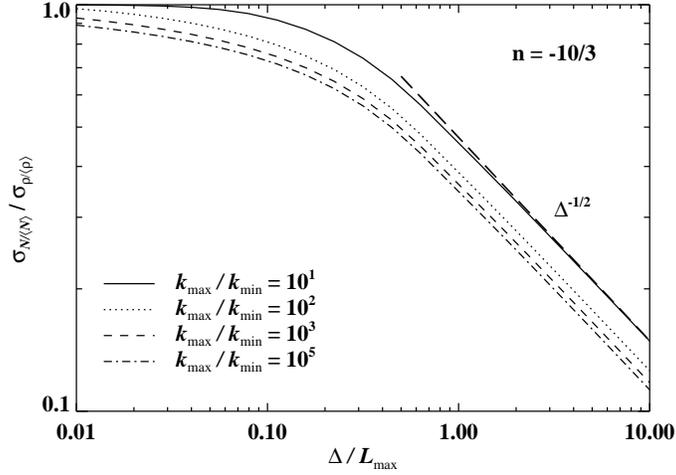}
  \caption{\label{sigma_N}
  Standard deviation of the column density, given by the ratio $\sigma_{N/\left<N\right>}/
  \sigma_{\rho/\left<\rho\right>}$, as function of slice thickness $\Delta/L_{\rm max}$. The power
  spectrum of the local density is taken to be Kolmogorov with $n=-10/3$. Also shown is the
  dependence on the dynamic range $k_{\rm max}/k_{\rm min}$ taken to be $10^1$, $10^2$, $10^3$, and $10^5$. The long dashed line shows the behaviour in the limit of thick slices
  where the thickness is larger then the maximum turbulent scale $L_{\rm max}$.}
\end{figure}

Equivalent to the variance of the local density the variance of the column density is given by the total power of the column densities:
\begin{equation}
  \sigma_{N}^2 = \frac{1}{(2\pi)^2}\int{\rm d} \mbox{\boldmath$K$}\,\tilde P(\mbox{\boldmath$K$}).
\end{equation}
In general the power spectrum of the column densities along the $z$-axis is determined by:
\begin{equation}
  \label{eqpowercol}
  \tilde P(\mbox{\boldmath$K$}) = \frac{1}{2\pi}\int{\rm d}k_{\rm z}\,\tilde P(\mbox{\boldmath$K$},k_{\rm z})\,\tilde W(k_{\rm z}\Delta)
\end{equation}
with $\tilde W(k_z\Delta)=|{W}(k_z\Delta)|^2$ where  $W(k_z\Delta)$
is the Fourier transform of the window function which defines the shape of the slice through the turbulent density structure \citep{Lazarian2000,Fischera2004a}. 
For a simple box-window, as used here, we have $W(k_z\Delta)=\Delta^2 \sin^2(k_z\Delta/2)/(k_z\Delta/2)^2$. 

We want to call a slice through the turbulent medium thick or thin if its thickness is either larger or smaller than the maximum scale length $L_{\rm max}$.
If the slice is thin with $\Delta\ll L_{\rm max}$ the power
spectrum of the column density is given by $\tilde P(K)\propto K^{n+1}$ for $1/L_{\rm max}< K \ll 1/\Delta$. On the other hand, if the thickness of the slice
is larger than the maximum 'cloud' size ($\Delta > L_{\rm max}$) the power spectrum
is well described by $\tilde P(K)\propto K^n$ and therefore by the same power law as the
power spectrum of the local density \citep{Lazarian2000,Fischera2004a}.

 In the limit of very thin slices smaller than the minimum `cloud' size
the fluctuations of the column densities are the ones of the local density. Due to averaging effects in the line of sights the fluctuations of the normalised column densities decrease with slice thickness (Fig.~\ref{sigma_N}). For slices thicker than the maximum cloud size the variation as function of slice thickness becomes a simple power law with $\sigma_{N/\left<N\right>}/\sigma_{\rho/\left<\rho\right>}\propto \Delta^{-1/2}$. The actual value of the standard deviation of the column density depends on the statistical properties. In a medium with a larger dynamic range the averaging effect in the line of sights is larger and the fluctuations of the column densities therefore weaker. 


\paragraph{The functional form of the PDF} 

\begin{figure}
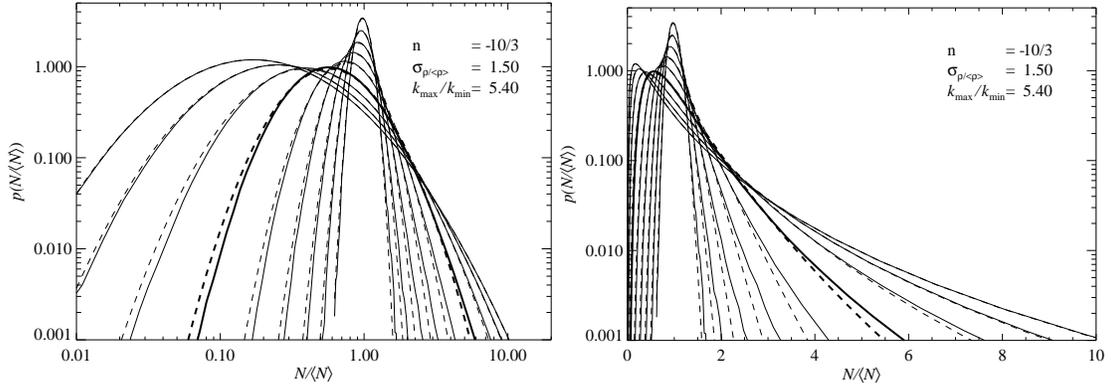

  \includegraphics[width=0.49\hsize]{fig2a.epsi}
  \includegraphics[width=0.49\hsize]{fig2b.epsi}
  \caption{\label{pdfcolumn}
  PDF of the column density $N/\left<N\right>$ through a simulated density structure with a 
  log-normal density
  distribution of the local density. The power spectrum of the local density is assumed to be
  Kolmogorov with $n=-10/3$ and the standard deviation of the local density 
  $\sigma_{\rho/\left<\rho\right>}=1.5$. The dynamic range is chosen to be
  $L_{\rm max}/L_{\rm min}=5.4$. The slice thickness through the density structure is
  taken to be}
\end{figure}

Based on MHD simulation of isothermal forced turbulence it has been found \citep{Ostriker2001} that the functional form of the PDF of the column density is approximately also log-normal. The
convergence of the log-normal density distribution of the column density into a Gaussian
density distribution in the limit of thick slices has been analysed in some detail by \citet{Vazquez2001}. They found by using density structures that the PDF is skewed in comparison to the ideal log-normal density distribution.

A more quantitative approach has been presented by \citet{Fischera2004a} based on simulated
density structures of an idealised isothermal turbulent medium with a log-normal density distribution of the local density and where the power is given by a defined power law.
The obtained distributions can be well approximated by a log-normal density distribution (Fig.~\ref{pdfcolumn}). However, deviations do exist as the fitted log-normal density distribution predicts too high probabilities at low but too low probabilities at high column densities. The deviation increases for broader distributions of the local densities. In addition the standard deviation of the fitted distributions is in general lower than the exact value \citep{Fischera2004a}. Despite these uncertainties it seems to be a first good approximation to simplify the functional form of the PDF by a simple log-normal density distribution where the variance is simply
determined by the thickness of the slice and the statistical properties of the local density.

\subsection{The Distant Foreground Screen}

We first consider a situation where the stars are seen through a turbulent dusty screen which is distant from the observer and the emitting stars. In this case the scattering makes only a small contribution to the collected star light and can be neglected. For simplicity we assume that the dust properties are not varying with local density or radiation field so that the optical depth is proportional to the column density. As the underlying extinction curve we use the mean extinction curve of our Galaxy which we adopted from Weingartner \& Draine \cite{Weingartner2001}.

At each wavelength the turbulent screen leads on average to an effective extinction given by 
\begin{equation}
  \label{eqtaueff}
   \tau_{\rm eff} =  -\ln\left(\int{\rm d}y\,p(y)\,e^{-e^y\left<\tau\right>}\right),
\end{equation}
where $p(y)$ is the normal density distribution and where $e^y =\tau/\left<\tau\right>$.


\begin{figure}
  \includegraphics[width=.6\textwidth]{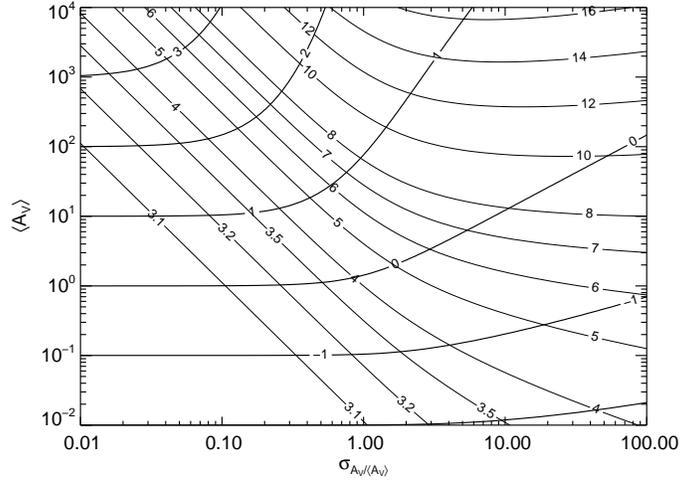}
  \caption{\label{parameterscreen}
	The parameter space of the distant turbulent screen given by
	the mean attenuation $\left<\rm A_V\right>$ and the standard 
	variation $\sigma_{\rm A_V/\left<A_V\right>}=\sigma_{N/\left<N\right>}$ 
	of the column density. The attenuation curves $\rm A_\lambda/A_V$ are
	well defined by the absolute to relative attenuation $\rm R_V^A$ shown 
	as thin solid lines. The actual attenuation $\rm A_V$ is shown as the thick 
	solid lines.
  }
\end{figure}

\begin{figure}
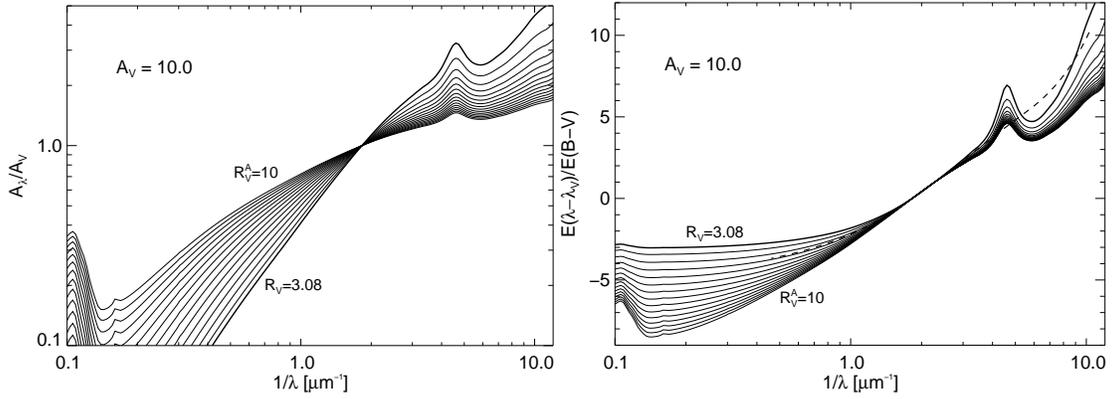

  \includegraphics[width=0.49\hsize]{fig4a.epsi}
  \includegraphics[width=0.49\hsize]{fig4b.epsi}
  \caption{\label{attenuationcurvesscreen}
	Attenuation curves $\rm A_\lambda/A_{V}$ and reddening curves
	$\rm E(\lambda-V)/E(B-V)$ of a distant foreground screen. The curves
	are well defined by the absolute to relative attenuation $\rm R_V^A$ which
	is varied from 3.5 to 10 in steps of $\Delta \rm R_V^A=0.5$. The thick line
	is the assumed extinction curve given by \citet{Weingartner2001}. The actual curves
	do not strongly depend on the absolute
	attenuation in V which is taken to be $\rm A_V=10$. The broken line is the 
	so called `Calzetti-extinction-curve'.
  }
\end{figure}

In general a turbulent screen is more transparent in comparison to a homogeneous screen and
the attenuation curve flatter than the underlying extinction curve. The effect of the turbulent screen on the attenuation increases with dust content or $\rm \left<A_V\right>$ and wider distributions of the column density (Fig.~\ref{parameterscreen}). 
The flattening can be characterised by the absolute to relative attenuation $\rm R_V^A=A_V/E(B-V)$ which is higher in case of flatter attenuation curves. In case of the foreground screen the $\rm R_V^A$-value is furthermore a perfect parameter to determine the curvature of the attenuation curves (Fig.~\ref{attenuationcurvesscreen}) over the whole wavelength range \cite{Fischera2005a}.\footnote{In general the curvature depends also on the geometry so that attenuation curves of different geometries but the same $\rm R_V^A$-value might deviate.} 
The dependence on the mean attenuation $\rm \left<A_V\right>$ on the other hand is only weak.
In the limit of small fluctuations with $\sigma_{\ln A_V}<1$ the relative attenuation $\rm A_\lambda/\left<A_\lambda\right>$ and the curvature depend only on the product $\sigma_{\ln A_V}^2\left<\rm A_V\right>$ \citep{Fischera2005a} so that in highly optically thick media also small fluctuations can lead to flatter attenuation curves.

\subsubsection{The `Calzetti Extinction Law' }

For the nearby universe we have some knowledge about the dust obscuration. The so called `Calzetti-extinction-law' \citep{Calzetti2001} has been obtained from a huge sample of star-burst galaxies and can be used to derive the intrinsic spectrum of those galaxies. The curve is flatter in comparison to the mean extinction curve in our galaxy and also flatter than
the ones derived for the Large and the Small Magellanic Cloud.
In addition the feature at 2200~\AA{} which is very prominent in case of the mean extinction curve of the Milky Way seems to be absent. The absence is generally explained by the destruction of the carriers of this feature by the higher UV-radiation field in those galaxies. The flatter curvature in the optical is thought to be related to the clumpy medium in front of the stars.

\begin{figure}
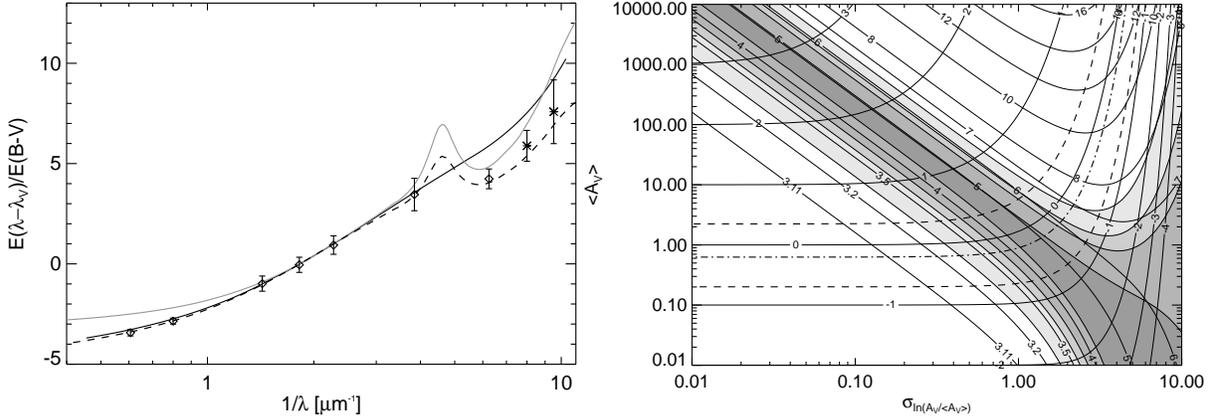

  \includegraphics[height=0.25\textheight]{fig5a.epsi} 
  \includegraphics[height=0.25\textheight]{fig5b.epsi}
	\caption{\label{fitcalzetti}
	Result of the $\chi^2$-fit to the `Calzetti extinction curve'. The 'Calzetti-curve' is shown as thick solid line in the left hand figure. Also shown is the mean extinction curve of our
	galaxy (\citet{Weingartner2001}) used in the model. The fit has been applied to the data point which are based on Calzetti \cite{Calzetti1997} and Leitherer et al. \cite{Leitherer2002}. The best fit is shown as dashed line. Regions of confidence are given as filled contours (68, 90, 95, 99\%) on the right hand figure in the parameter plane defined by $\rm A_V$ and $\rm \sigma_{\rm A_V/\left<A_V\right>}$. The dashed lines show the range of $\rm A_V$ determined for star-burst galaxies. The dashed-dotted line is the mean value.}
\end{figure}

The overall curvature of the Calzetti-Curve can be naturally explained by a turbulent foreground screen \cite{Fischera2003}. The $\chi^2$-fit based on photometric measurements from star-burst galaxies is excellent (even though the minimum $\chi^2$ of 0.048 suggest an overestimate of the errors) (Fig.~\ref{fitcalzetti}). The fit provides a most likely $\rm R_V^A$-value of $4.75\pm 0.45$ which is in excellent agreement with the original value $\rm R_V^A=4.88\pm 0.98$ derived for star-bursts \cite{Calzetti1997}. Based on IR-measurements this value has been corrected to $\rm R_V^A=4.05\pm 0.80$. The agreement with our value of a turbulent screen, however, is still good.

The measured range of $\rm A_V$ of star-burst galaxies provides a standard deviation of the 
optical depth (column density) in the range $0.66$ to $11.2$. If we apply the relation between Mach-number and standard deviation we obtain a minimum Mach number in the range 1.3 to 22. 

It might be useful to compare those numbers with the expectation of the cold neutral medium (CNM) and the warm neutral medium (WNM) in our own galaxy. The velocity dispersion for the CNM and for the WNM has been measured (based on HI and CO observation, respectively) to be  6-8 km/s and 7-10 km/s. Assuming a temperature of 100 and 6000 K for the CNM and the WNM the Mach number is roughly 12 and 1.8. If we use the relation \ref{densmachcorr} the expected range for $\sigma_{\rho/\left<\rho\right>}$ is in the range 0.9 to 6 consistent with the Mach numbers found by fitting the Calzetti curve by a simple turbulent screen.

\subsubsection{The Effect of Slice Thickness on the Attenuation Curve}

For a better understanding of a turbulent density structure on the attenuation it is
important to know how the attenuation changes with the thickness of the dust layers in front
of the stars. As an application one might think of a simple geometry of disk-like galaxies where the stars are embedded in a turbulent dusty layer and therefore
seen through a turbulent screen. If the medium would be homogeneous the optical thickness of the dusty layers in front of the stars would increase with inclination angle as $1/\cos i$. 
The dependence is somewhat weaker if the medium is turbulent as the relative attenuation $\rm A_\lambda/\left<A_\lambda\right>$ decreases with the thickness of the screen. 
In addition the attenuation curves should flatten with viewing angle as the 
$\rm R_V^A$-value increases as a function of $\Delta/L_{\rm max}$. 


\begin{figure}
  \includegraphics[width=0.7\hsize]{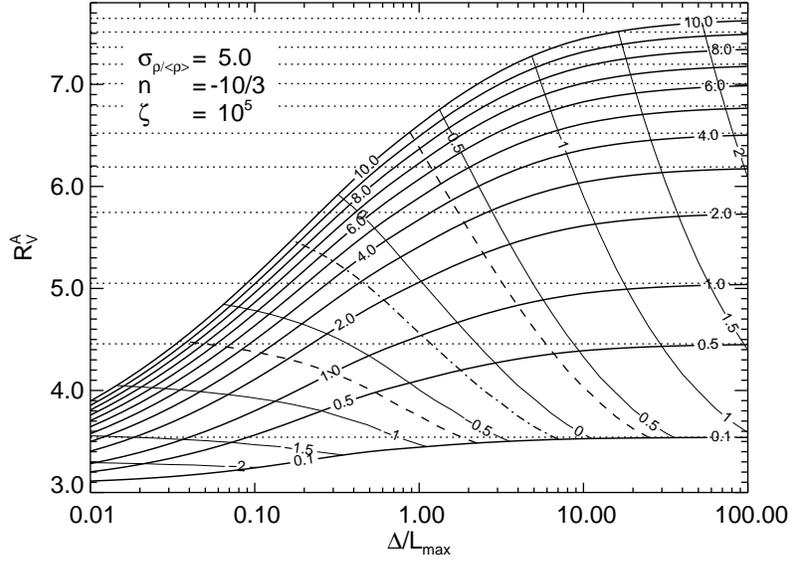}
  \caption{\label{rv_thickness}
  Variation of the absolute to relative attenuation $\rm R_V^A$
  	as function of the slice thickness $\bigtriangleup/L_{\rm max}$ for a certain
	choice of the turbulent density structure. 
	The thick solid lines refer to turbulent media with different mean attenuation
	$\left<\rm A_V\right>_{L_{\rm max}}$ varied from 0.1 to 10. The absolute
	attenuation $\rm A_V$ is shown as thin solid lines. The absolute $\rm R_V^A$-values
	in the limit of thick slices are given as dotted lines. Also shown are the maximum,
	the mean, and the minimum attenuation $\rm A_V$, derived for star-burst galaxies 
	(dashed, dashed-dotted, and dashed line).  
	}
\end{figure}

In the limit of thick slices (or high inclination angles $i$) the relative attenuation $\rm A_\lambda/\left<A_\lambda\right>$ and the absolute to relative attenuation $\rm R_V^A$ approach
asymptotically a minimum and a maximum value, respectively. The limiting values
are, for a given power spectrum and dynamic range, only determined by the product $\left<\tau_{\rm V}\right>_{L_{\rm max}}\sigma^2_{\rho/\left<\rho\right>}$ \citep{Fischera2005a}.


\subsubsection{The Relation between $\rm R_V^A$ and $\rm A_V$}

The relation between the two observable quantities $\rm A_V$ and $\rm R_V^A$ has consequences for the interpretation
of the change of an intrinsic spectral energy distribution from galaxies caused by dust obscuration.  A stellar spectrum well described over a certain wavelength range by a simple power
law $I_\lambda\propto \lambda^\beta$ with power $\beta$ \citep{Calzetti1994} will have a different
functional form if seen through a dusty screen. 
If the spectrum is observed at two different wavelengths $\lambda_1$ and $\lambda_2$ the 
inferred power $\tilde \beta$ would be different from $\beta$ by
\begin{equation}
  \Delta\beta = \beta-\tilde \beta=\frac{\tau_{\lambda_1}-\tau_{\lambda_2}}{\ln\lambda_1-\ln\lambda_2} =  \frac{1}{\rm \tilde R_{\lambda_2}^{\rm A}}\frac{\tau_{\lambda_2}}{\ln\lambda_1-\ln\lambda_2}.
\end{equation}
Here, we introduced the absolute to relative attenuation 
$\rm \tilde R_{\lambda_2}^{\rm A}=\tau_{\lambda_2}/(\tau_{\lambda_1}-\tau_{\lambda_2})$. $\tau_{\lambda_1}$ and $\tau_{\lambda_2}$ are the effective optical depths at wavelength $\lambda_1$ and $\lambda_2$.


Due to turbulence the variation of $\Delta \beta$ is in general not directly proportional to $\rm A_\lambda$. As $\rm R_{\lambda_2}^A\ge R_{\lambda_2}$ the change of the spectral index $\beta$ should be furthermore less strong in comparison to a homogeneous screen.

\subsection{The Non-Scattering Turbulent Slab}

To demonstrate the effect of the geometry on the attenuation curve we consider the extreme case that the stars are homogeneously distributed throughout the turbulent slab. For simplicity the scattering is ignored so that the dust layers in front of the stars at each depth can be approximated by a distant foreground screen. 

\begin{figure}
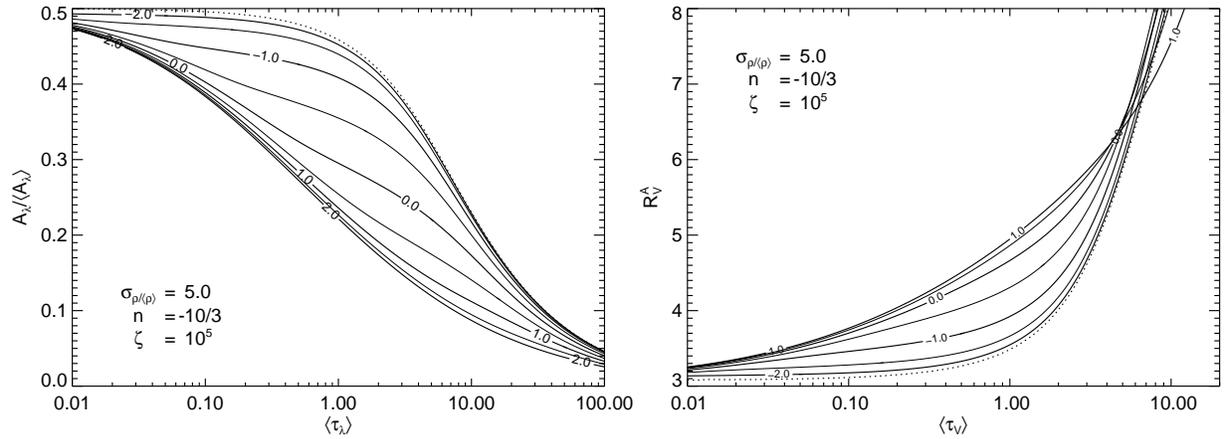

  \includegraphics[height=0.26\textheight]{fig7a.epsi}
  \includegraphics[height=0.26\textheight]{fig7b.epsi}
  \caption{\label{slabavrv}
	Relative attenuation $\rm A_V/\left<A_V\right>$ and absolute to relative attenuation
	$\rm R_V^A$ as function of mean optical depth $\left<\tau\right>$ through a turbulent
	slab in which the sources are homogeneously distributed. 
	The dotted curves show the behaviour in case of a homogeneous medium.
	The solid lines correspond to different assumptions of the mean optical depth 
	at one maximum scale $L_{\rm max}$. The labels refer to 
	$\log\left<\tau\right>_{L_{\rm max}}$.
  }
\end{figure}

The slab-geometry leads even in case of the homogeneous medium  to a lower attenuation and to a flatter attenuation curve (Fig.~\ref{slabavrv}). 
In the limit of an optically thin
and optically thick limit the effective optical depth is $\tilde \tau = 0.5\left<\tau\right>$ and 
$\tilde\tau = \ln\left<\tau\right>$, respectively. The $\rm R_V^A$-value increases strongly when the slab becomes optically thick. In the limit of an optically thick slab the absolute to relative attenuation varies as $\rm R_V^A = \ln\left<\tau\right>/\ln(1+1/\rm R_V)$ and therefore
is directly proportional to the effective optical depth $\tilde \tau$.



By introducing a turbulent density structure the medium becomes more transparent
and the relative attenuation decreases in comparison to the homogeneous slab (Fig.~\ref{slabavrv}). In the region $\left<\tau\right><\sim 4$ the turbulence furthermore leads to a flatter attenuation curve. The behaviour at higher mean optical depth depends on the dust content per maximum scale length. In rather optically thick environments the attenuation curve can be slightly steeper than the very flat attenuation curve expected for a homogeneous medium.



\section{Conclusion}

The described model relates the dust attenuation to the physical properties of the 
turbulent ISM and therefore might be a solution to obtain accurate corrections important
to measure the star-formation rate against redshift. As shown, a distant turbulent screen can naturally explain the attenuation curve derived for star-burst galaxies and may be also applicable
for other star-forming regions. 

The idealised model of a turbulent density structure might also be helpful to understand the radiative transport through the turbulent ISM in the disks of normal galaxies. The model may furthermore provide, if used in a full radiative transfer code which also includes the thermal emission from dust grains, a better understanding of the SED of star-burst galaxies or HII-regions.


\begin{theacknowledgments}
	The author would like to thank the Research School of Astronomy and Astrophysics
	and the Australian National University for their support. I would like to
	acknowledge financial support for this research through the ARC Discovery project
	DP0208445. Finally I would like to thank the Max-Planck-Institut f\"ur Kernphysik where
	I was a guest while I was writing this proceedings paper.
\end{theacknowledgments}


\bibliographystyle{aipproc}   

\bibliography{reference}

\IfFileExists{\jobname.bbl}{}
 {\typeout{}
  \typeout{******************************************}
  \typeout{** Please run "bibtex \jobname" to optain}
  \typeout{** the bibliography and then re-run LaTeX}
  \typeout{** twice to fix the references!}
  \typeout{******************************************}
  \typeout{}
 }

\end{document}